\documentclass[aps, showpacs, twocolumn]{revtex4}
\usepackage{amsmath,amssymb}

\usepackage{tabularx}

\usepackage{epsfig}
\usepackage{graphicx}

\begin{document}

\title{Assisted chaotic inflation in brane-world cosmology}

\author{Grigoris Panotopoulos\footnote{grigoris@theorie.physik.uni-muenchen.de}}

\date{\today}


\address{ASC, Department of Physics LMU, Theresienstr. 37, 80333 Munich, Germany}


\begin{abstract}
Assisted chaotic inflation in brane cosmology is discussed. We work in the
framework of Randall-Sundrum (RS) II model, in which adopting the RS condition
the only parameter is the five-dimensional Planck mass. Using the scalar
spectral index and the amplitude of scalar perturbations we determine both the
mass of the scalar fields responsible for inflation and the fundamental Planck
mass of the higher-dimensional theory. We find that the mass of the scalars
has the typical value of the inflaton mass in chaotic inflation
($M_I \sim 10^{13}$~GeV) and that the five-dimensional Planck mass is very
close to the GUT (Grand Unified Theories) scale
($M_5 \sim (10^{16}-10^{17})$~GeV). Furthermore, no matter how many scalar 
fields we use it is not possible to have chaotic inflation with field 
values below the fundamental Planck mass.
\end{abstract}

\pacs{98.80.-k, 98.80.Es, 98.80.Cq, 04.50.+h}

\maketitle

Inflation~\cite{inflation} has become the standard paradigm for the early
Universe, because it solves some outstanding problems present in the standard
Hot Big-Bang cosmology, like the flatness and horizon problems, the problem of
unwanted relics, such as magnetic monopoles, and produces the cosmological
fluctuations for the formation of the structure that we observe today. The
recent spectacular CMB data from the WMAP satellite~\cite{wmap} have
strengthen the inflationary idea, since the observations indicate an
\emph{almost} scale-free spectrum of Gaussian adiabatic density fluctuations,
just as predicted by simple models of inflation. According to chaotic
inflation with a potential for the inflaton field $\phi$ of the form
$V=(1/2) m^2 \phi^2$, the WMAP normalization condition requires for the
inflaton mass $m$ that $m=1.8 \times 10^{13} \: GeV$~\cite{Ellis:2003sq}.
However, a yet unsolved problem about inflation is that we do not know how to
integrate it with ideas in particle physics. For example, we would like to
identify the inflaton, the scalar field that drives inflation, with one of the
known fields of particle physics. Furthermore, it is important that the
inflaton potential emerges naturally from an underline fundamental theory. See however~\cite{Allahverdi:2006iq} as an example of realistic embedding withing particle physics. In this model the inflaton has Standard Model gauge charges instead of being a ad-hoc gauge singlet added by hand.

It is known~\cite{powerlawinflation} that if there is just a single scalar
field $\phi$, the exponential potential
$V(\phi)=V_0 \: \textrm{exp}(-\lambda \phi/m_{pl})$,where $m_{pl}$ is Planck mass and
$\lambda=\sqrt{16 \pi/p}$, leads to
power-law solution $a(t) \sim t^p$, with $a(t)$ the scale factor of the
universe as a function of the cosmic time. For steep potentials ($p<1$) the
solution is decelerating and therefore it does not correspond to inflation,
while for shallow potentials ($p>1$) the solution is accelerating and thus it
is an inflationary solution. Scalar potentials of exponential form emerge in
supergravity theories, however these theories typically predict steep
exponential potentials. In assisted inflation~\cite{assistedinflation} many
scalar fields drive inflation in a cooperative way and so it is possible to
obtain inflationary solutions even in such models.

Over the last years the brane-world models have been attracting a lot of
attention as a novel higher-dimensional theory. Brane models are inspired from
M/string theory and although they are not yet derivable from the fundamental
theory, at least they contain the basic ingredients, like extra dimensions,
higher-dimensional objects (branes), higher-curvature corrections to gravity
(Gauss-Bonnet) etc. Since string theory claims to give us a fundamental
description of nature it is important to study what kind of cosmology it
predicts. Furthermore, despite the fact that inflationary models have been
analyzed in standard four-dimensional cosmology, it is challenging to discuss
them in alternative gravitational theories as well. In brane-world models it
is assumed that the standard model particles are confined on a 3-brane while
gravity resides in the whole higher dimensional spacetime. The model first
proposed by Randall and Sundrum (RSII)~\cite{rs}, is a simple and interesting
one, and its cosmological evolutions have been intensively
investigated. An incomplete list can be seen e.g. in~\cite{langlois}. In the present
work we would like to study
assisted chaotic inflation in the framework of RSII model. According to that
model, our 4-dimensional
universe is realized on the 3-brane with a positive tension located at the UV
boundary of 5-dimensional AdS spacetime. In the bulk there is just a
cosmological constant $\Lambda_{5}$, whereas on the brane there is matter with
energy-momentum tensor $\tau_{\mu \nu}$. Also, the five dimensional Planck
mass is denoted by $M_{5}$ and the brane tension is denoted by $T$.

We start by presenting the theoretical framework in which we will be working, namely the RSII brane model and the dynamics of assisted inflation, as well as the two quantities (spectral index and amplitude of perturbations) that will link the theoretical model to observations.

If Einstein's equations hold in the five dimensional bulk, then it has been shown in \cite{shiromizu} that the effective four-dimensional Einstein's equations induced on the brane can be written as
\begin{equation}
G_{\mu \nu}+\Lambda_{4} g_{\mu \nu}=\frac{8 \pi}{m_{pl}^2} \tau_{\mu \nu}+(\frac{1}{M_{5}^3})^2 \pi_{\mu \nu}-E_{\mu \nu}
\end{equation}
where $g_{\mu \nu}$ is the induced metric on the brane, $\pi_{\mu \nu}=\frac{1}{12} \: \tau \: \tau_{\mu \nu}+\frac{1}{8} \: g_{\mu \nu} \: \tau_{\alpha \beta} \: \tau^{\alpha \beta}-\frac{1}{4} \: \tau_{\mu \alpha} \: \tau_{\nu}^{\alpha}-\frac{1}{24} \: \tau^2 \: g_{\mu \nu}$, $\Lambda_{4}$ is the effective four-dimensional cosmological constant, $m_{pl}$ is the usual four-dimensional Planck mass and $E_{\mu \nu} \equiv C_{\beta \rho \sigma} ^\alpha \: n_{\alpha} \: n^{\rho} \: g_{\mu} ^{\beta} \: g_{\nu} ^{\sigma}$ is a projection of the five-dimensional Weyl tensor $C_{\alpha \beta \rho \sigma}$, where $n^{\alpha}$ is the unit vector normal to the brane.
The tensors $\pi_{\mu \nu}$ and $E_{\mu \nu}$ describe the influence of the bulk in brane dynamics. The five-dimensional quantities are related to the corresponding four-dimensional ones through the relations
\begin{equation}
m_{pl}=4 \: \sqrt{\frac{3 \pi}{T}} \: M_{5}^3
\end{equation}
and
\begin{equation}
\Lambda_{4}=\frac{1}{2 M_{5}^3} \left( \Lambda_{5}+\frac{T^2}{6 M_{5}^3} \right )
\end{equation}
In a cosmological model in which the induced metric on the brane $g_{\mu \nu}$ has the form of  a spatially flat Friedmann-Robertson-Walker model, with scale factor $a(t)$, the Friedmann-like equation on the brane has the generalized form~\cite{langlois}
\begin{equation}
H^2=\frac{\Lambda_{4}}{3}+\frac{8 \pi}{3 m_{pl}^2}  \rho+\frac{1}{36 M_{5}^6} \rho^2+\frac{C}{a^4}
\end{equation}
where $C$ is an integration constant arising from $E_{\mu \nu}$. The cosmological constant term and the term linear in $\rho$ are familiar from the four-dimensional conventional cosmology. The extra terms, i.e the ``dark radiation'' term and the term quadratic in $\rho$, are there because of the presence of the extra dimension. Adopting the Randall-Sundrum fine-tuning
\begin{equation}
\Lambda_{5}=-\frac{T^2}{6 M_{5}^3}
\end{equation}
the four-dimensional cosmological constant vanishes. In addition, the dark radiation term will be quickly diluted during inflation and therefore we shall neglect it. So the generalized Friedmann equation takes the final form
\begin{equation}
H^2=\frac{8 \pi G}{3} \rho \left (1+\frac{\rho}{\rho_0} \right )
\end{equation}
where
\begin{equation}
\rho_0=96 \pi G M_{5}^6
\end{equation}
with $G$ the Newton's constant. One can see that the evolution of the early universe can be divided into two eras. In the low-energy regime $\rho \ll \rho_0$ the first term dominates and we recover the usual Friedmann equation of the conventional four-dimensional cosmology. In the high-energy regime $\rho_0 \ll \rho$ the second term dominates and we get an unconventional expansion law for the universe. Since inflation is assumed to take place in the high-energy regime we shall keep only the term which is quadratic in $\rho$ and from now on we shall make use of the new law for expansion of the universe
\begin{equation}
H(\rho)=\frac{\rho}{6 M_5^3}
\end{equation}

Assisted inflation model~\cite{assistedinflation} involves more than one
scalar fields $\phi_i$ $(i=1,2,\cdots,p)$. Let us present here the dynamics of this model for chaotic inflation, i.e. a
simple quadratic
potential $V(\phi)=m^2\phi^2/2$. We consider the case in which all scalars have the
same potential (i.e. same mass). The scalar fields are taken to be
noninteracting
but  are minimally coupled to
gravity. Under these assumptions the total scalar potential is $W(\phi_1, \cdots, \phi_p)=\sum_{i=1}^p V(\phi_i)$ and the equations of motion are
\begin{eqnarray}
\ddot \phi_i & = & -{m^2 \phi_i}-3 H \dot \phi_i \\
H & = & \frac{1}{6 M_5^3}\sum_{i=1}^p \left(V(\phi_i)+ {\dot
\phi_i^2\over2}\right)
\end{eqnarray}
To simplify things we shall consider a particular solution in which all fields are taken equal
$\phi_1=\phi_2=\cdots=\phi_p=\Phi$. The simplified equations in the slow-roll approximation become
\begin{eqnarray}
0 & = & -{m^2 \Phi}-3 H \dot \Phi \\
H & = & \frac{1}{6 M_5^3} \: p \: V(\Phi)
\end{eqnarray}

The amplitude of scalar perturbations is defined by
\begin{equation}
A_s=\frac{2}{5} \: \mathcal{P}_{\mathcal{R}}^{1/2}
\end{equation}
where $\mathcal{P}_{\mathcal{R}}$ is the spectrum of the curvature perturbation given by~\cite{stewart}
\begin{equation}
\mathcal{P}_{\mathcal{R}}=\left(\frac{H}{2 \pi} \right)^2 \: \delta_{ij} \: \frac{\partial N}{\partial \phi_i} \: \frac{\partial N}{\partial \phi_j}
\end{equation}
with $N$ the number of e-folds remaining (defined by $N=-\int dt H$), which satisfies the useful equation
\begin{equation}
\sum_{i=1}^p \frac{\partial N}{\partial \phi_i} \: \dot{\phi}_i = -H
\end{equation}
Furthermore, the spectral index is given by~\cite{stewart}
\begin{equation}
n-1=2\frac{\dot{H}}{H^2}-2 \frac{\frac{\partial N}{\partial \phi_i} \: \frac{\partial N}{\partial \phi_j}(\frac{8 \pi \: \dot{\phi}_i \: \dot{\phi}_j}{m_{pl}^2 H^2}-\frac{W_{,i,j}}{3 H^2})}{\delta_{ij} \: \frac{\partial N}{\partial \phi_i} \: \frac{\partial N}{\partial \phi_j}}
\end{equation}
In our case in which $H(\rho)=\rho/(6 M_5^3)$, $V(\phi)=m^2 \phi^2/2$ and all fields are taken to be equal we obtain
\begin{eqnarray}
\delta_{ij} \: \frac{\partial N}{\partial \phi_i} \: \frac{\partial N}{\partial \phi_j} & = & \frac{H^2}{p \: \dot{\Phi}^2} \\
\frac{\partial N}{\partial \phi_i} \: \frac{\partial N}{\partial \phi_j} \: \dot{\phi}_i \: \dot{\phi}_j & = & H^2 \\
W_{,i,j} & = & m^2 \: \delta_{i,j} \\
N(\Phi) & = & \frac{p \: m^2 \: \Phi^4}{192 M_5^6}
\end{eqnarray}

We fix the number of e-folds to be $N_*=60$ and allow for the number $p$ of the scalar fields to take values $p=2,10,50,100$. In our model there are just two free parameters, namely the five-dimensional Planck mass $M_5$ and the mass $m$ of the scalar fields. We take for $A_s=2 \times 10^{-5}$ and $n=0.95$. From these two observational facts we can determine both $M_5$ and $m$. First, from the amplitude $A_s$ we determine the ratio $m/M_5$ and then from the spectral index we compute $M_5$. We use the formulae
\begin{eqnarray}
\frac{m}{M_5} & = & \left(\frac{15 \pi A_s}{p^{5/4} \: (192 N_*)^{1/4}} \right)^{2/3} \\
n & = & 1-\frac{1}{2 \: p \: N_*}-\sqrt{\frac{p}{3 N_*}} \: \frac{24 \pi}{p^2} \: \left(\frac{M_5}{m_{pl}} \right)^2 \: \frac{M_5}{m}
\end{eqnarray}
Our results can be shown below
\begin{itemize}
\item For $p=2$, \quad $M_5=6.24 \times 10^{16}~\textrm{GeV}, m=7.08 \times 10^{13}~\textrm{GeV}$
\item For $p=10$, \quad $M_5=1.11 \times 10^{17}~\textrm{GeV}, m=3.30 \times 10^{13}~\textrm{GeV}$
\item For $p=50$, \quad $M_5=1.90 \times 10^{17}~\textrm{GeV}, m=1.48 \times 10^{13}~\textrm{GeV}$
\item For $p=100$, \quad $M_5=2.40 \times 10^{17}~\textrm{GeV}, m=1.05 \times 10^{13}~\textrm{GeV}$
\end{itemize}

One can see that the mass $m$ is of the order $\sim 10^{13}$~GeV, which is
the typical inflaton mass in chaotic inflation with just one scalar field.
In addition, we see that the fundamental Planck mass is close to the GUT
scale, $M_{GUT} \sim 10^{16}$~GeV. Assisted inflation typically involves
a large number of scalar fields, for which we obtain $M_5 \sim 10^{17}$~GeV.

In the case of standard four-dimensional cosmology, assisted inflation is used in 
order to have chaotic inflation with sub-Planckian field values. The main motivation 
for assisted inflation in a brane-world scenario would be to have inflation with field 
values below the fundamental Planck mass. We now compute the minimum number of scalar 
fields for which $\Phi_* < M_5$ (if possible). We make use of the formula (another way 
of writing equation (20) for $N_*=60$)
\begin{equation}
\frac{\Phi_*}{M_5}=\left ( \frac{60 \times 192}{p} \right )^{1/4} \: \left ( \frac{M_5}{m} \right )^{1/2}
\end{equation}
Then using equation (21) we obtain that $\Phi_*/M_5$ goes like $p^{1/6}$ and it 
is always larger than unity. Therefore it is impossible to have chaotic inflation 
with field values below $M_5$.

To summarize, in the present work we have discussed assisted chaotic inflation in
the RSII brane-world scenario. In this model the universe expands according to a
novel Friedmann-like equation and inflation is driven by $p$ non-interacting scalar
fields which are minimally coupled to gravity. For simplicity we have considered the
case in which all fields are taken to be equal with a common quadratic potential
$V(\phi)=m^2 \phi^2/2$. Using the observational facts that
$A_s=2 \times 10^{-5}, n=0.95$ we have determined the two free parameters of the
model, i.e. the mass $m$ of the scalar fields and the five-dimensional Planck
mass $M_5$. Our results show that $m \sim 10^{13}$~GeV, a typical inflaton mass
in chaotic inflation with just a single scalar field, and that
$M_5 \sim (10^{16}-10^{17})$~GeV. Furthermore, we have found that it not possible 
to have chaotic inflation with field values below the fundamental Planck mass.

\section*{Acknowlegements}

We would like to thank T.~N.~Tomaras for reading the manuscript and for useful comments. This work was supported by project ``Particle Cosmology''.


\begin{references}
\bibitem{inflation}  A.~H.~Guth,
   ``The Inflationary Universe: A Possible Solution To The Horizon And Flatness
  Problems,''
  Phys.\ Rev.\ D {\bf 23} (1981) 347; \\
D.~H.~Lyth and A.~Riotto,
   ``Particle physics models of inflation and the cosmological density
  perturbation,''
  Phys.\ Rept.\  {\bf 314} (1999) 1
  [arXiv:hep-ph/9807278]; \\
 N.~Straumann,
   ``From primordial quantum fluctuations to the anisotropies of the cosmic
  microwave background radiation,''
  Annalen Phys.\  {\bf 15} (2006) 701
  [arXiv:hep-ph/0505249].
\bibitem{wmap}  D.~N.~Spergel {\it et al.},
   ``Wilkinson Microwave Anisotropy Probe (WMAP) three year results:
  Implications for cosmology,''
  arXiv:astro-ph/0603449.
\bibitem{Ellis:2003sq} J.~R.~Ellis, M.~Raidal and T.~Yanagida,
   ``Sneutrino inflation in the light of WMAP: Reheating, leptogenesis and
  flavor-violating lepton decays,''
  Phys.\ Lett.\ B {\bf 581} (2004) 9
  [arXiv:hep-ph/0303242].
\bibitem{Allahverdi:2006iq} R.~Allahverdi, K.~Enqvist, J.~Garcia-Bellido and A.~Mazumdar,
  Phys.\ Rev.\ Lett.\  {\bf 97} (2006) 191304
  [arXiv:hep-ph/0605035].
\bibitem{powerlawinflation} F.~Lucchin and S.~Matarrese,
  ``Power Law Inflation,''
  Phys.\ Rev.\ D {\bf 32} (1985) 1316.
\bibitem{assistedinflation} A.~R.~Liddle, A.~Mazumdar and F.~E.~Schunck,
  ``Assisted inflation,''
  Phys.\ Rev.\ D {\bf 58} (1998) 061301
  [arXiv:astro-ph/9804177].
\bibitem{rs} L.Randall and R.Sundrum,
``An alternative to compactification,''
Phys.\ Rev.\ Lett.\  {\bf 83} (1999) 4690
[arXiv:hep-th/9906064].
\bibitem{langlois} P.Binetruy, C.Deffayet and D.Langlois,
``Non-conventional cosmology from a brane-universe,''
Nucl.\ Phys.\ B {\bf 565} (2000) 269
[arXiv:hep-th/9905012]. \\
P.Binetruy, C.Deffayet, U.Ellwanger and D.Langlois,
``Brane cosmological evolution in a bulk with cosmological constant,''
Phys.\ Lett.\ B {\bf 477} (2000) 285
[arXiv:hep-th/9910219]. \\
C.~Deffayet, G.~R.~Dvali and G.~Gabadadze,
  ``Accelerated universe from gravity leaking to extra dimensions,''
  Phys.\ Rev.\ D {\bf 65} (2002) 044023
  [arXiv:astro-ph/0105068]; \\
E.Kiritsis, N.Tetradis and T.N.Tomaras,
``Induced gravity on RS branes,''
JHEP {\bf 0203} (2002) 019
[arXiv:hep-th/0202037]. \\
E.Kiritsis, G.Kofinas, N.Tetradis, T.N.Tomaras and V.Zarikas,
``Cosmological evolution with brane-bulk energy exchange,''
JHEP {\bf 0302} (2003) 035
[arXiv:hep-th/0207060]; \\
R.~A.~Brown, R.~Maartens, E.~Papantonopoulos and V.~Zamarias,
  ``A late-accelerating universe with no dark energy - and no big bang,''
  JCAP {\bf 0511} (2005) 008
  [arXiv:gr-qc/0508116]; \\
G.~Kofinas, G.~Panotopoulos and T.~N.~Tomaras,
  ``Brane-bulk energy exchange: A model with the present universe as a  global
  attractor,''
  JHEP {\bf 0601} (2006) 107
  [arXiv:hep-th/0510207]; \\
 G.~Panotopoulos,
  ``Sneutrino inflation in Gauss-Bonnet brane-world cosmology, the  gravitino
  problem and leptogenesis,''
  Nucl.\ Phys.\ B {\bf 745} (2006) 49
  [arXiv:hep-ph/0511040].
\bibitem{shiromizu} T.Shiromizu, K.i.Maeda and M.Sasaki,
``The Einstein equations on the 3-brane world,''
Phys.\ Rev.\ D {\bf 62} (2000) 024012
[arXiv:gr-qc/9910076]; \\
A.~N.~Aliev and A.~E.~Gumrukcuoglu,
  ``Gravitational field equations on and off a 3-brane world,''
  Class.\ Quant.\ Grav.\  {\bf 21} (2004) 5081
  [arXiv:hep-th/0407095].
\bibitem{stewart}  M.~Sasaki and E.~D.~Stewart,
   ``A General Analytic Formula For The Spectral Index Of The Density
  Perturbations Produced During Inflation,''
  Prog.\ Theor.\ Phys.\  {\bf 95} (1996) 71
  [arXiv:astro-ph/9507001].
\end{references}
\end{document}